\documentclass[prb,twocolumn,showkeys,showpacs,preprintnumbers,amssymb,amsmath,superscriptaddress]{revtex4}
\usepackage{graphicx}

\begin{document}

\title{Theoretical study of kinks on screw dislocation in silicon}

\date{\today}

\author{L. Pizzagalli}
\email{Laurent.Pizzagalli@univ-poitiers.fr}

\affiliation{Laboratoire de M\'etallurgie Physique, CNRS UMR 6630, Universit\'e
de Poitiers,  B.P. 30179, 86962 Futuroscope Chasseneuil Cedex, France}

\author{A. Pedersen}

\affiliation{Science Institute, University of Iceland, Dunhaga 3, IS-107 Reykjavik, Iceland}

\author{A. Arnaldsson}

\affiliation{Science Institute, University of Iceland, Dunhaga 3, IS-107 Reykjavik, Iceland}

\author{H. J{\'o}nsson}

\affiliation{Science Institute, University of Iceland, Dunhaga 3, IS-107 Reykjavik, Iceland}

\author{P. Beauchamp}

\affiliation{Laboratoire de M\'etallurgie Physique, CNRS UMR 6630, Universit\'e
de Poitiers,  B.P. 30179, 86962 Futuroscope Chasseneuil Cedex, France}

\begin{abstract}

Theoretical calculations of the structure, formation and migration
of kinks on a non-dissociated screw dislocation in silicon have been carried out using density functional theory calculations as well as calculations based on interatomic potential functions. The results show that the structure of a single kink is characterized by a narrow core and highly stretched bonds between some of the atoms. The formation energy of a single kink ranges from 0.9 to 1.36~eV, and is of the same order
as that for kinks on partial dislocations. However, the kinks migrate almost freely along the line of an undissociated dislocation unlike what is found for partial dislocations. The effect of stress has also been investigated in order to compare with previous silicon deformation experiments which have been carried out at low temperature and high stress. The energy barrier associated with the formation of a stable kink pair becomes as low as 0.65~eV for an applied stress 
on the order of 1~GPa, indicating that displacements of screw dislocations likely occur via thermally activated formation of kink pairs at room temperature. 

\end{abstract}



\preprint{report number}

\maketitle

\section{Introduction}

New information about the plasticity of silicon and other diamond lattice materials has recently emerged. 
Measurements at low temperature have shown deformation without failure when silicon samples are submitted to large shear stress.\cite{Rab00JPCM,Rab00PSS,Rab01SM,Asa05MSE} 
These experiments have been carried out either by applying a high confining pressure or by doing scratch tests. The observed configurations of dislocations
differ strongly from the dissociated
dislocations seen after deformation at higher temperature.\cite{Due91SSMS} 
The low temperature dislocations are
undissociated and most likely belong to and glide along the widely spaced (111) ``shuffle planes''.\cite{Hir82WIL}  
They mainly align along several favored orientations,
$<\!110\!>$ (screw dislocation), $<\!112\!>$ ($30^\circ$ dislocation) and $<\!123\!>$ ($41^\circ$ dislocation). Similarly, in III-V compounds with the zinc-blende structure, deformation
experiments at low temperature under high confining pressure indicate that the low temperature
plastic deformation is governed by undissociated screw dislocations.\cite{Suz98PML,Suz99PMA}  
To understand the plastic properties of these materials, it is then important to characterize these undissociated dislocations.

Most theoretical studies of dislocations in diamond lattice materials have focused on dissociated dislocations with partials in the glide set (the narrowly spaced (111) planes). 
For silicon, simulations have clearly shown that 
the 30$^\circ$ partial dislocation core is ($2\times1$) reconstructed along the dislocation line.\cite{Due91PRB,Leh99EMIS} 
Two core structures have been proposed for
the 90$^\circ$ partial dislocation. They are close in energy\cite{Big92PRL,Val98PRL} and both appear to be stable under some conditions.\cite{Leh98PRL,Mir03PRB}  
Partial dislocations have also been theoretically investigated in other materials, such as Ge,\cite{Nun98PRB,Nun00JPCM}  diamond,\cite{Nun98PRB,Bla00PRL,Blu02PRB,Blu03JPCM,Blu03PRB} GaAs\cite{Bec02JPCM,Jus02JPCM} and other III-V compounds,\cite{Jus00JPCM} as well as in silicon carbide.\cite{Blu02JPCM,Blu03PRB2} 
It is generally believed that due to large Peierls barriers, the partial dislocations in semiconductors move by formation of kink pairs and subsequent migration of kinks.\cite{Hir82WIL} 
In this framework, the formation energy of a single kink, $F_k(90^\circ)=0.73$~eV and $F_k(30^\circ)=0.80$~eV, and kink migration energy $W_m=1.24$~eV have experimentally been determined.\cite{Kol96PRL} 
Theoretical studies of this are difficult since many possible kink configurations have to be considered\cite{Nun00JPCM,Bul95PMA,Bul01PMA} and the calculations need to be carried out for large systems in order to reduce the effect of the boundaries. 
This makes first principles calculations where electronic degrees of freedom are included particularly challenging and less reliable methods based on interatomic potential functions or tight-binding are usually employed. As a result, calculated values of the formation energy that have been reported range from 0.04~eV to 1.2~eV ($F_k(90^\circ)$) and from 0.25~eV to 2.15~eV ($F_k(30^\circ)$), and values for the migration energy range from 0.6~eV to 1.8~eV ($W_m(90^\circ)$) and from 0.7~eV to 2.1~eV ($W_m(30^\circ)$).\cite{Leh99EMIS,Bul95PMA,Bul01PMA,Hua95PRL,Obe95PRB,Val98PRL} 

Much less information is available for undissociated dislocations, mainly because low temperature experiments require high stress conditions. Naturally, previous investigations have focussed on the screw orientation because, among all characteristics, it always plays a special role. Indeed, it allows for cross-slip, and in the particular case of the diamond cubic structure, it is the orientation where the transition between the glide set and the shuffle set is possible without any structural rearrangements relying on diffusion. Moreover, in observations after low-temperature deformation, the screw appears as one of the favored orientations, what could indicate significant Peierls valleys.\cite{Rab01SM} Using first-principles methods, Arias and Joannopoulos\cite{Ari94PRL} have confirmed the stability of the screw dislocation placed at the centre of an hexagon, that is at the intersection of two \{111\} shuffle planes (Fig.~\ref{Kinkgeometry}). Pizzagalli et al. have shown\cite{Piz03PMA} that a screw centered at A is more stable than two other possible positions, B at the intersection of a glide and a shuffle plane and C at the intersection of two glide planes. The Peierls stress of the undissociated screw dislocation at position A has been determined, using ab initio simulations, to $0.07\mu$ ($\mu$ is the shear modulus of silicon).\cite{Piz04PML} Therefore, the undissociated screw can move under the effect of an applied stress, contrary to the partials in the glide set which cannot move without thermal activation. Allowing for period-doubling reconstruction of the undissociated screw dislocation core, Wang et al. have recently shown that the glide set C core has a lower energy than the shuffle A core.\cite{Wan06APL} This is certainly an important information to keep in mind when examining possible transitions from the shuffle set to the glide set. However, transmission electron microscopy observations of the dislocations after deformation at low temperature show that all parts of an expanding loop, having different characters, are undissociated.\cite{Rab00JPCM} Since all parts of a gliding loop necessarily glide in the same plane, the whole dislocation loop should belong to and glide in the shuffle set. Undissociated perfect dislocations have also been investigated in other materials such as SiC,\cite{Blu03PRB2,Piz05EPL} Ge,\cite{Piz05EPL} diamond,\cite{Blu02PRB,Blu03PRB,Piz05EPL} and GaN.\cite{Bel05PSS} 

There is a serious lack of knowledge about the mobility properties of undissociated dislocations, especially about the role of kinks, their structures and energetics. As far as we know, the only published work focuses on the relation between dislocation velocity and kinks and does not provide reliable data about kink energies.\cite{Koi05MSE} A determination of these quantities would improve the current knowledge of (i) the thermally-assisted  motion of the screw dislocation, through formation and expansion of kink pairs, and of (ii) the gradual transition for strictly screw character to non screw orientations within the expanding loop. From a more general perspective, Rabier and Demenet\cite{Rab00PSS} pointed out 
that the observation of undissociated dislocations fits into the analysis of Duesbery and Joos,\cite{Due96PML} which, if extrapolated, predicts a transition from dissociated dislocations in the glide set in the high temperature/low stress domain, to undissociated dislocations in the shuffle set in the low temperature/large stress domain. A possible explanation could be related to differences in mobility properties between dissociated and undissociated dislocations, as a function of temperature and stress. Another explanation could be the different formation/multiplication properties. Since partial dislocations mobility is already well characterized, determining the mobility of undissociated dislocations would enable one to confirm the first hypothesis.  

In this work, we have investigated the mobility properties of an undissociated screw dislocation in the shuffle set, by determining kinks structures and energies from calculations using both an interatomic potential and first principles. The paper is organized as follows. First, models, methods and analysis techniques are described. Then the structure and the migration and formation energies together with the effect of an applied stress are determined using an interatomic potential and reported. In the third part, we describe the results obtained from first principles calculations. Finally, we discuss our results in relation with experiments and known properties of the mobility of dissociated dislocations in silicon.

\section{Methods and models}

\subsection{Calculations methods}

\begin{figure}
\begin{center}
\includegraphics*[width=8.6cm]{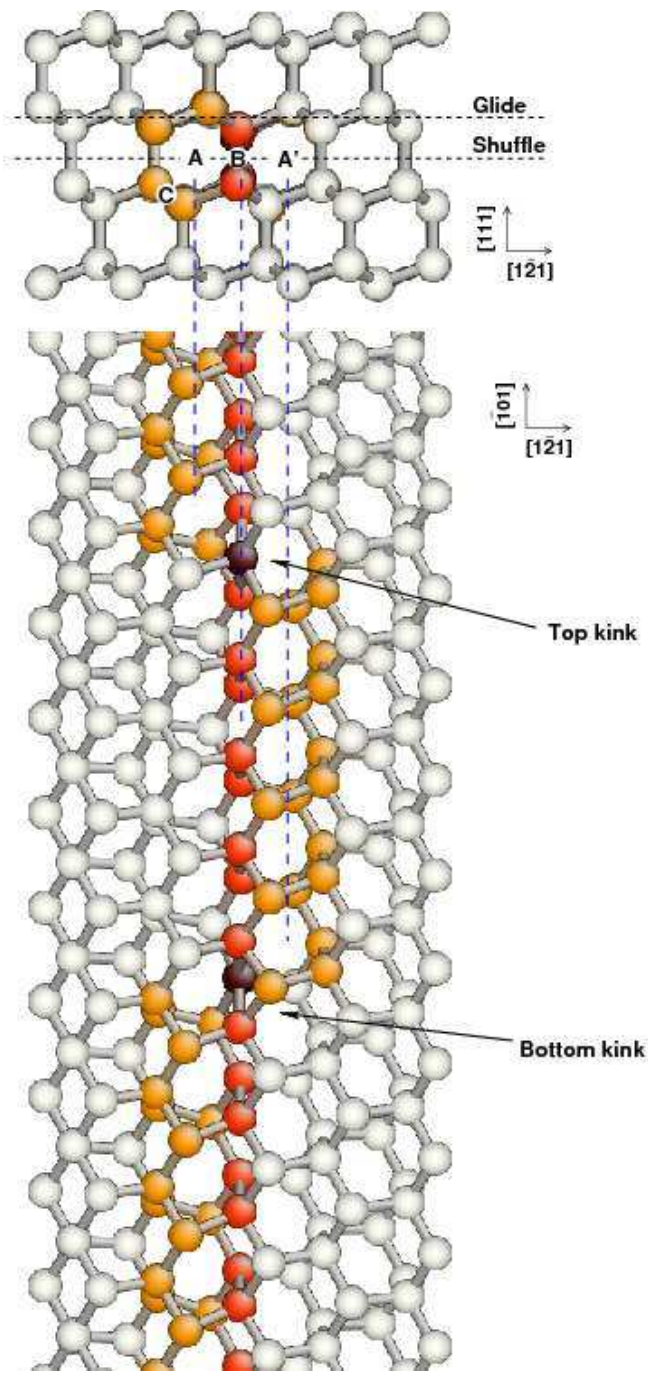}
\caption{(color online) Ball-stick top and side representations of the system used in EDIP calculations (only atoms in the center of the cell are shown). The initial screw dislocation line, oriented along $[\bar{1}01]$, is located at position A in the shuffle plane. The screw dislocation segment between the formed kinks is located at position A', also in the shuffle plane. B (mixed shuffle-glide) and C (glide) are other possible positions for the dislocation line. Orange spheres show atoms in the immediate vicinity of the dislocation core, belonging to hexagons around positions A or A'. Red spheres are atoms belonging to the dimer row along $[\bar{1}01]$  centered on position B, between the two screw dislocations locations. Black spheres are atoms in the center of kinks with a coordination 5.} \label{Kinkgeometry}
\end{center}
\end{figure}

The modelling of silicon has been done with both a semi-empirical interatomic potential and a first principles description. Among the available potentials for silicon, we have selected the EDIP potential,\cite{Baz97PRB} because it has been specifically designed for describing extended defects,\cite{Jus98PRB} and it adequately reproduces the stability of the screw dislocation.\cite{Piz03PMA} With this potential, the lattice constant is equal to the experimental data, i.e. $a=5.43$~\AA.
First principles calculations have been performed in the framework of Density Functional Theory (DFT),\cite{Hoh64PR,Koh65PR} using the VASP distribution.\cite{Kre96PRB,Kre96CMS} Ionic interactions have been described with a Si ultrasoft pseudopotential,\cite{Van90PRB} and wave functions have been expanded on a plane waves basis with an energy cutoff of 140~eV. The exchange-correlation contributions have been modelled by the PW91 Generalized Gradient Approximation (GGA).\cite{Per91AKA} Only the $\Gamma$-point has been considered for the Brillouin zone sampling. Finally, the system was considered fully relaxed when all forces were below $3\times10^{-3}$~eV~\AA$^{-1}$. With these parameters, a lattice constant  $a=5.475$~\AA\ and a bulk modulus of 99.7~GPa have been obtained, in good agreement with experimental data.

Minimum energy paths (MEP) associated with the creation and migration of kinks have been determined using the Nudged Elastic Band (NEB) technique.\cite{Jon98WS} Investigations of the formation and migration of kinks in body-centered-cubic or L1$_2$ materials have already been done with a similar approach.\cite{Wen00AM,Nga01PRL,Nga04CMS} Both improved tangent and climbing images algorithms\cite{Hen00JCP,Hen00JCPb} have been employed in this work. Generally, we found that migration and formation mechanisms were simple enough to be described with a small number of images. Nevertheless, we used as many as 30~images in EDIP simulations, and 9 in first principles calculations. 

\subsection{Model}~\label{model}

EDIP potential calculations were done using a large parallelepipedic box, oriented along $\mathbf{i}=1/4[1\bar{2}1]$ (\^X), $\mathbf{j}=1/6[111]$ (\^Y), and $\mathbf{k}=1/2[\bar{1}01]$ (\^Z). Typical system dimensions are $38\mathbf{i}\times78\mathbf{j}\times20\mathbf{k}$, i.e. 126.357\,\AA\,$\times$\,122.265\,\AA\,$\times$\,76.792\,\AA, accounting for 59280 silicon atoms. Vacuum was added next to the \^X and \^Y surfaces. A kinked screw dislocation was introduced in the center of the cell according to the following procedure: (1) along \^Z the system is split up into 20 slices of width $\mathbf{k}$ (2) for each slice, the displacements field due to a screw dislocation of Burgers vector $\mathbf{b}=\mathbf{k}$, calculated with anisotropic elasticity theory, is applied on all atoms (3) for slices on the outside (called the A-slice in the following), the screw dislocation is located at the shuffle position A, whereas for slices in the cell center (the A'-slices), it is located one hexagon away along \^X, at position A' (Fig.~1). This procedure allows to generate initial configurations with two opposite kinks along the screw dislocation line, with a kink-kink separation varying by steps of $\mathbf{b}$. 

Specific boundary conditions have been used in the EDIP calculations. The two surfaces having \^X as normal have been left free to relax, since we checked that the system size is large enough to have negligible surface relaxation. Conversely, along \^Y, the surfaces have been frozen in order to apply a well defined shear stress on the system by rigidly shifting the \^Y surfaces. Along \^Z, periodic conditions have been applied, causing our model to represent an infinite number of interacting kinks. In one specific case the \^Z surfaces were frozen in order to analyse one isolated kink.  

DFT numerical simulations have been performed using cluster-like systems with periodic boundary conditions in all directions. A vacuum of 5~\AA\ is surrounding the clusters, in order to prevent spurious image interactions. The input structures for the DFT simulations have been cut from relaxed EDIP calculations, thus retaining the initial crystal orientation. All surface atoms have been saturated by hydrogens, and kept at their initial positions, to preserve the long-range strain field as obtained by EDIP calculations.  For the study of kink migration, the systems dimensions were approximately $5\mathbf{i}\times5\mathbf{j}\times6\mathbf{k}$, consisting of 336~Si and 197~H atoms. For kink formation, we have considered larger systems including 528-529~Si and 269-271~H atoms, depending on the configuration, with dimensions about $5\mathbf{i}\times5\mathbf{j}\times9\mathbf{k}$.

\subsection{Kink analysis}\label{kinkanalysis}

\begin{figure}
\begin{center}
\includegraphics*[width=8.6cm]{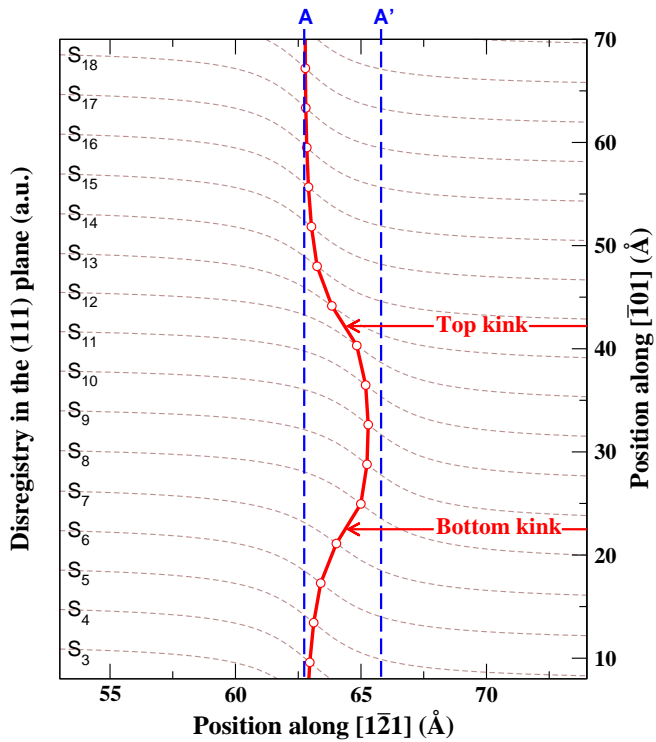}
\caption{(color online) Kinks positions along \^Z$=[\bar{1}01]$ obtained from the variation of the screw dislocation core center. Thin dashed lines represent the disregistry in the shuffle plane for the ``slice'' $S_i$ (magnified by a factor 3 for clarity). Circles on the red thick line show the position of the screw dislocation core along \^X$=[1\bar{2}1]$ for each ``slice'' $S_i$, ranging from A to A' (see Fig.~1). From this curve, the kinks position along \^Z$=[\bar{1}01]$ can be determined.} \label{Kinkanalysis}
\end{center}
\end{figure}

In order to investigate properties as a function of the kink-kink separation, the position of both kinks along \^Z must be known with a good accuracy. As will be shown in section~\ref{EDIP}, the stable structure of a kink is characterized by a defined symmetry, that is why the geometric position for a kink can be defined as a local symmetry center in the structure. In the following, this approach is called local determination. 

However, local determination is not suited for kink migration analysis, since the symmetry brakes when the structure is evolving. Instead, here, we propose to use fits based on elasticity theory in order to determine the position of kinks. As in the previous section, we assume that the system can be divided along \^Z into slices of width $\mathbf{b}=\mathbf{k}$. For each slice, the position of the screw dislocation core is obtained from the disregistry along $[\bar{1}01]$ between atoms on either sides of the initial shuffle $(111)$ plane (Fig.~2). The variation of the disregistry $\delta z_i$ along \^X for the slice $S_i$ is then fitted by the following expression

\begin{equation}
\delta z_i=b\left[-\frac{1}{\pi}\arctan\left(\frac{x-x^c_i}{\Delta_i}\right)+\frac{1}{2}\right],
\end{equation}

yielding for each slice $S_i$ the position of the screw dislocation core $x^c_i$ along \^X, and a quantity $\Delta_i$ proportional to the width of the dislocation core. 

Computing the variation of the dislocation position $x^c$ along \^Z is then straightforward, and we found that such a variation can be accurately described using two $\arctan$ functions, one for each kink. Practically, we considered the following fitting expression: 

\begin{equation}
x^c=\frac{h}{\pi}\left[\arctan\left(\frac{x-x^k_1}{\delta^k_1}\right)-\arctan\left(\frac{x-x^k_2}{\delta^k_2}\right)\right]+x^c_A,
\label{equ:kinkWidth}\end{equation}

where $x^k_1$, $x^k_2$, $\delta^k_1$, $\delta^k_2$ are positions and widths of both kinks, and $x^c_A$ is the position A of the screw dislocation (see Fig.~1), and $h=a\sqrt{6}/4=3.325$~\AA\ is the kink height. We refer to this procedure as a global determination in the next sections. It allows to estimate the kinks position even during their migration, since it does not rely on the core structure of the kink, but rather on the modification of the strain field due to the kink. Moreover, the width of the kinks are also obtained with this analysis. It has to be noted that when a shear strain is applied on the system, the calculated disregistry has been corrected accordingly.

\section{EDIP results}\label{EDIP}

\subsection{Single kink}\label{singlekink}

\begin{figure}
\begin{center}
\includegraphics*[width=8.6cm]{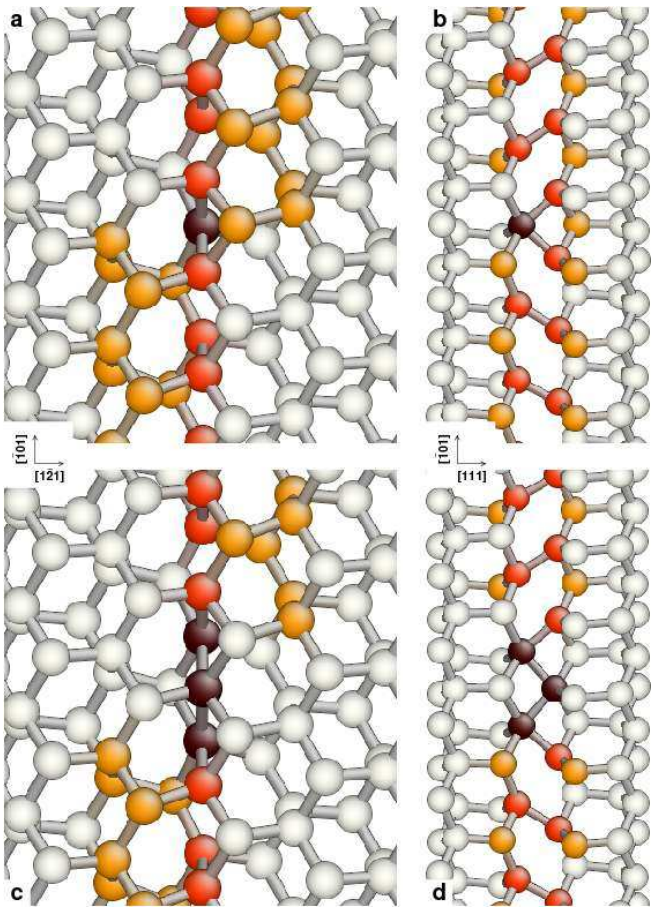}
\caption{(color online) Ball-stick representations of two kink structures stable in EDIP calculations (top: narrow kink, bottom: wide kink), along two different orientations, (111) (left) and $(1\bar{2}1)$ (right) projections. Same color convention than in figure~\ref{Kinkgeometry}.} \label{EDIPkinkstruct}
\end{center}
\end{figure}

We first describe the results obtained with the EDIP potential. The initial configurations, when searching for the stable kink structure, have been generated using the procedure described in the section~\ref{model}, i.e. by introducing elastic displacements corresponding to a kinked screw dislocation. Immediate relaxation leads to a metastable kink structure containing at least one dangling bond. Simulated annealing or by-hand displacement of few atoms yield a more stable reconstructed kink structure, shown in the figure~\ref{EDIPkinkstruct}a,b. It includes a 5-coordinated atom in the core of the kink, and can be understood by examining the \^Z=$[\bar{1}01]$ stacking of Si dimers (colored in red in figure~\ref{EDIPkinkstruct}b). These dimers are tilted relative to \^Z, and the tilt direction tells whether the dimer belongs to an A-slice (positive angle) or A'-slice (negative angle). The 5-coordinated atom serves as the linking core where dimers from A to A' connect. The two linking bonds are characterized by a length of 2.53~\AA, and tilting angle of $\pm42.4^\circ$ relatively to $(\bar{1}01)$. For comparison equivalent dimers in a straight screw dislocation have a bond length of 2.44~\AA\ and a tilting angle of 21$^\circ$ (not shown), showing that the bonds in the linking core are more severely stretched. Obviously, as shown  in the figure~\ref{Kinkgeometry}, top and bottom kink configurations are symmetric, as they can be transformed in each other with a C$_2$ rotation along the \^Z axis at position B. For reasons that will become apparent below, we name this configuration the narrow kink. 

Several attempts to find other stable kink structures have been made. For instance, we have tried to position the center of the kink in the middle of a slice  (as defined in section~\ref{model}), rather than between two slices. No satisfactory results were obtained. However, another kink structure, depicted in figure~\ref{EDIPkinkstruct}c,d emerged from NEB calculations of the barrier for kink migration. It is alike the kink structure described above, but instead of one, three 5-coordinated atoms constitute the kink center. Bonds in the center have length of 2.56~\AA\ and form an angle of 48.4$^\circ$, whereas bonds linking a 5-coordinated atom with a 4-coordinated atom have length of 2.52~\AA\ and form an angle of 44.2$^\circ$ with previous bonds. We name this second configuration the wide kink. 

Using concepts already proposed for silicon partials,\cite{Bul01PMA} it is possible to consider the wide kink structure as the dissociation of the narrow kink into partial kinks. These partial kinks are then separated by a one dimensional stacking-fault, which can be identified as a screw dislocation located between A and A', in a mixed shuffle/glide position B (see figure~\ref{Kinkgeometry}). We have previously shown that such a screw dislocation is stable with EDIP, but not with first principles computations.\cite{Piz03PMA} In a first approximation, according to isotropic elasticity theory,\cite{Hir82WIL} the kink dissociation results in an energy change 

\begin{equation}
\Delta E=\frac{\mu b^{2}(h/2)^{2}}{8\pi}\,\frac{(1+\nu)}{(1-\nu)}\,\frac{1}{d}\,+\,\gamma d,
\end{equation}

with the shear modulus $\mu=68.1$~GPa $=0.425$~eV~\AA$^{-3}$, the Poisson coefficient $\nu=0.218$, the partial kink height $h/2=a\sqrt{6}/8=1.66$~\AA, and $d$ being the partial kinks separation. The first term represents the elastic energy decrease when partial kinks separates, whereas the second term is the linear energy increase due to the 1D stacking fault formation. For the line energy $\gamma$, we use the EDIP calculated energy difference between a screw dislocation in position A and in the mixed position B,\cite{Piz03PMA} i.e. $\gamma=0.06$~eV~\AA$^{-1}$. Taking the derivative of the above formula, we found that the energy is minimal for a partial kink separation $d=4.2$~\AA $=1.1$~$b$. This value seems to be reasonable in comparison with the structure of both narrow and wide kinks. However, since the model is crude the result should only be used as an indication of the stability of wide kinks.

We have tried to determine the relative stability of both kink configurations, from single kink calculations made in a large cell using full fixed boundary conditions. Unfortunately, narrow and wide kinks have almost the same energy, the difference being only few meV. The calculation is likely troublesome due to the large fixed boundary area. Surface effects could add up and become too large when compared with the energy difference between both configurations. Also, along \^Z, both kinks are not exactly located at the same distance from the surfaces, due to their different structures. Nevertheless, we will show in the following that the energy difference is indeed very small, about 100~meV, but in favor of the narrow kink. 

\subsection{Migration}\label{migration}

\begin{figure}
\begin{center}
\includegraphics*[width=8.6cm]{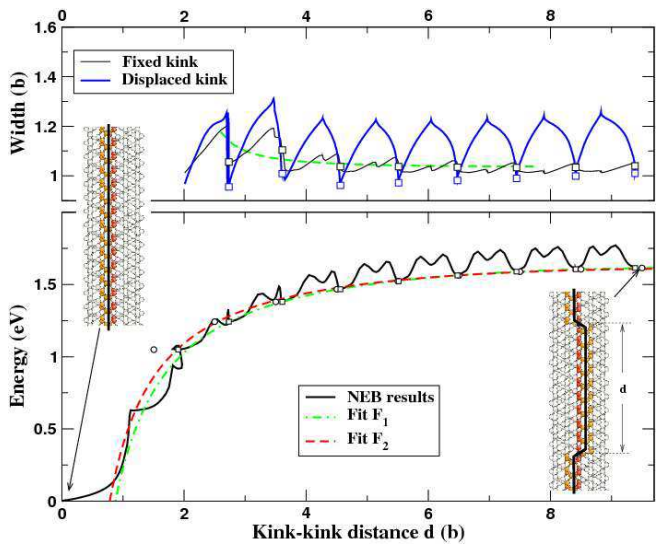}
\caption{(color online) Kink pair energy (bottom graph) and kinks width (top graph) versus kink-kink separation $d$ (in Burgers vector $b$), as obtained from NEB EDIP calculations. Insets represent the initial system (perfect screw) and the final one ($d\approx10b$). In the bottom graph, squares (circles) mark stable configurations with $d$ calculated with the kink position global (local) determination, respectively. These squares (circles) have been fitted using elasticity theory. In the top graph, thick (thin) lines show the kink width variation for the displaced (fixed) kink, while squares mark the position of stable configurations for varying $d$. } \label{EDIPcurve}
\end{center}
\end{figure}

We have investigated the full process of a kink pair formation and the subsequent migration of one kink by performing a series of NEB calculation, with the system described in the section~\ref{model}. The initial NEB configuration is a perfect screw, whereas in the final image, formed kinks are separated by about $10b$ (Fig.~\ref{EDIPcurve}). We performed a series of 10 consecutive NEB calculations. In each calculation a band of 30 images was used connecting an initial configuration containing two narrow kinks separated by $d=nb$ (${n=0}$ equals perfect screw) and a final configuration where $d$ is increased to $(n+1)b$. For each relaxed image, the position of the kinks have been determined using the global method described in the section~\ref{model}. 

\begin{figure*}
\begin{center}
\includegraphics*[width=16cm]{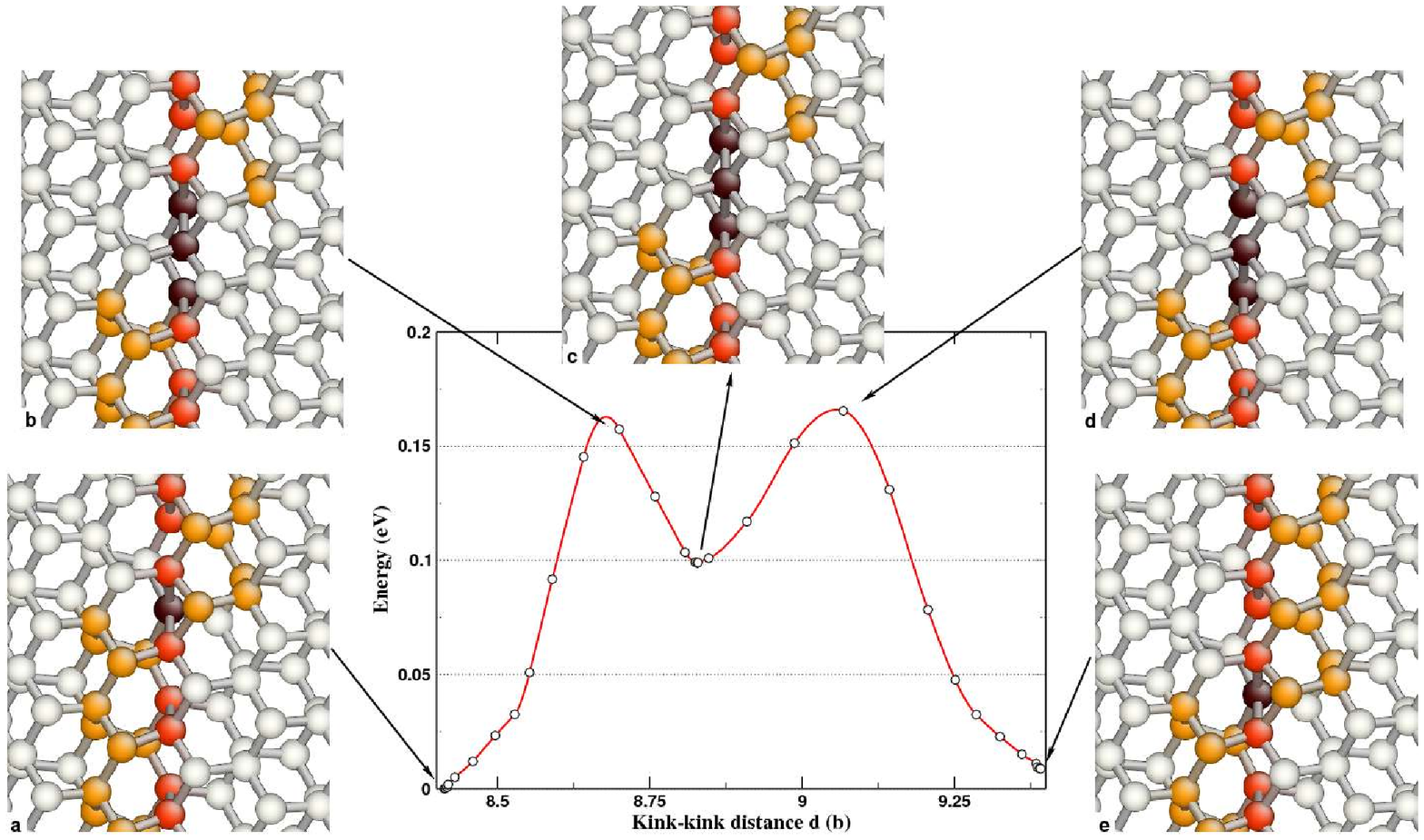}
\caption{(color online) Kink pair energy versus kink-kink separation $d$ (in Burgers vector $b$). NEB calculated points (white circles) are linked together with a cubic spline as a guide for the eyes. Selected kink configurations are also shown, with the same color convention than in figure~\ref{Kinkgeometry}.} \label{EDIPmigration}
\end{center}
\end{figure*}

The minimal energy path (MEP) mapped to kink-kink distance is shown in figure~\ref{EDIPcurve}. Relatively to the perfect screw, the energy steeply increases until $d\sim2b$ is reached, it then bends of and reach a plateau value at $d=10b$, where the model, due to periodic boundary conditions, represents an infinite distribution of equally-separated kinks. From $d\sim4b$ to $d\sim10b$, a well defined repeated pattern with a period of $b$ clearly appears, what corresponds to the minimum energy path for the migration of a single kink. Figure~\ref{EDIPmigration} shows both the energy variation and the structural changes for a kink migrating from an initial configuration where the two narrow kinks are separated by $d\sim8.4b$ to the final state where $d\sim9.4b$. The metastable state emerging halfway through the migration process corresponds to a wide kink. This result suggests that (1) the wide kink is an intermediate state allowing the transition of a narrow kink from one stable position to another (2) the wide kink is less stable than the narrow one, with an energy difference of about 0.1~eV. 

The transition mechanism requires the formation and the breaking of only one bond. The initial structure with a 5-coordinated atom is progressively stretched along 
$[\bar{1}01]$, until a bond is formed with the next dimer (see inset a,b in figure~\ref{EDIPmigration} and figure~\ref{EDIPkinkstruct}). The kink structure containing three 5-coordinated atoms relaxes and the wide kink is obtained. To reach the final stable state the configuration evolves further after the breaking of the original bond, leading to a transition into a narrow kink configuration displaced by $b$ (Figure~\ref{EDIPmigration}). The kink width as determined with expression~\ref{equ:kinkWidth} is varying in full agreement with the above description. In fact, the width increases almost linearly from the initial narrow kink to the intermediate wide kink, then decreases similarly to the final narrow kink (Figure~\ref{EDIPcurve}). The determined widths are approximately $1.0b$ for a narrow kink and $1.2b$ for a wide kink, in good agreement with the simple model developed in the section~\ref{singlekink}. 

The kink migration process is associated with two energy barriers to overcome, the migration energy $W_m$ being the highest of these barriers. It can be determined  by substracting the kink-kink elastic interaction contribution, described in the next section, from the energy variation represented in the figure~\ref{EDIPcurve}. For kink-kink distances larger than $5b$, we found that the remaining energy variation associated with migration is well defined, with an average value  $W_m=158\pm5$~meV.

\subsection{Formation}

\begin{figure}
\begin{center}
\includegraphics*[width=8.6cm]{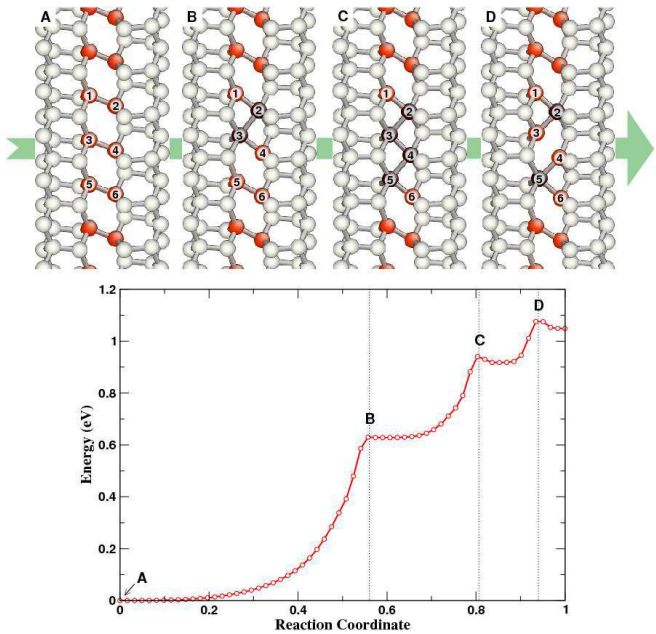}
\caption{(color online) Ball-stick representations ($(1\bar{2}1)$ projection) of the successive structures leading to a kink pair formation on a straight screw dislocation (top) and NEB calculated excess energy versus reaction coordinate (bottom). Dashed lines mark the energies associated with configurations on top. Same color convention than in figure~\ref{Kinkgeometry}.} \label{EDIPformation}
\end{center}
\end{figure}

Now we focus on the kink pair formation mechanism and the associated energy barrier. According to anisotropic elasticity theory,\cite{Hir82WIL} the energy variation as a function of the kink-kink distance $d$ is the elastic interaction between two opposite kinks and twice the formation energy, $F_k$, of a single kink,

\begin{equation}
\Delta E=-\frac{K}{d}+2\,F_k. \label{eqnformigr}
\end{equation}

In the isotropic approximation, $K$ is proportional to the shear modulus and is equal to $\mu b^{2}h^{2}(1+\nu)/[8\pi(1-\nu)]=1.12$~eV~b$=4.3$~eV.\AA. We used  expression~\ref{eqnformigr} to fit\cite{fit} the calculated energies of configurations corresponding to stable kinks, with kink positions determined with the global approach (Figure~\ref{EDIPcurve}). The best agreement is obtained for the values $K=1.6$~eV~b and $F_k=0.91$~eV, yielding the fitted curve labelled F$_1$. The full energy variation is well reproduced by this fit, except for short kink-kink separations for which elasticity theory is not applicable. A comparable curve F$_2$ was obtained when the local approach is used for determining kink positions instead, with $K=1.4$~eV~b and $F_k=0.90$~eV. It is noteworthy that very close values for $F_k$ are obtained for the two different kink position determination methods. Conversely, the elastic factor $K$ is rather sensitive to the energy variation for small $d$, for which kinks positions as determined with the two different approachs tend to differ. 

In the figure~\ref{EDIPformation}, the MEP and the structure of several intermediate configurations during the creation of a kink pair on a straight screw dislocation are shown. For the MEP, the reaction coordinate is used instead of $d$ in the energy plot, since the kink position analysis (section~\ref{model}) fails when kinks are not fully formed. The initial configuration is a straight screw dislocation located in position A, for which the Si dimers are stacked along $[\bar{1}01]$, centered on position B and all tilted in the same direction as shown in Fig.~\ref{EDIPformation}A. The kink formation begins with an increase in the angle of dimers 1-2 and 3-4, until atoms 2 and 3 are close enough to form a bond (Fig.~\ref{EDIPformation}B). This first reorganization is characterized by an exponential-like increase of the energy. The energy variation then shows a short plateau followed by another exponential-like increase. This corresponds to yet another process, in which the dimer 5-6 is progressively tilted, leading to the formation of a bond between atoms 4 and 5 (Fig.~\ref{EDIPformation}C). Finally, a third atomistic mechanism occurs, characterized by a small lowering of energy followed by a sharp increase. The bond between atoms 3 and 4 breaks causing two very close narrow kinks to be created (Fig.~\ref{EDIPformation}D). Other mechanisms for kink pair formation, involving the formation of wide kinks in particular, have been investigated and seen to result in paths being higher in energy. 

In the section~\ref{migration}, we have described the kink width variation for migration when the kinks are well separated. As soon as the kink-kink distance $d$ is larger than about 4$b$, the average widths for both kinks seem to be constant (Figure~\ref{EDIPcurve}). It is also intructive to examine the kinks widths variation when they are strongly interacting, even if for very small separations, it is difficult to draw any meaningful conclusions. Using the simple elastic model developed in the section~\ref{singlekink}, we have considered two interacting kinks separated by $d$, each one being dissociated in two partial kinks separated by a variable width. This model enable us to determine the kink width corresponding to the minimum energy for each of the values of $d$. The best solution, obtained for a line energy $\gamma$ of 0.06~eV~\AA$^{-1}$, is plotted in the Figure~\ref{EDIPcurve}. The agreement is not satisfying when the kinks are close, what either could be due to the elastic model is too simple, or that the kink width determination is not accurate enough for small $d$.

\subsection{Effect of stress}

The effect of stress on formation and migration of the kinks has been investigated by imposing a shear strain on the system. The shear strain $\epsilon_{ZY}$ has been chosen, since it is optimal for displacing the screw dislocation along \^X. Stress and strain are related through the appropriate elastic constant, here the $\langle\bar{1}01\rangle{111}$ shear modulus $G=1/3(C_{11}-C_{12}+C_{44})$ ($G=61.3$~GPa$=0.383$~eV~\AA$^{-3}$ for the EDIP potential). NEB calculations, similar to those reported in the previous sections, have been performed for shear strain values of 0.5\%, 1\% and 1.5\%. 

\begin{table*}[!]
\renewcommand{\arraystretch}{1.4}
\caption{Kink pair formation and migration data, from EDIP calculations and with kinks position determined using either the global or local determinations (see section~\ref{kinkanalysis}). The first two columns show the applied strain and the corresponding stress. $K$ is the elastic prefactor, $F_k$ the single kink formation energy, $\sigma$ the stress relaxed by kinks formation and migration, $W_m$ the kink migration energy, $E^*$ the energy barrier for a kink pair formation, and $d^*$ the minimum kink-kink distance for creating a stable kink pair.} \label{stresstable}
\begin{ruledtabular}
\begin{tabular}{ccccccccccc}
 & & \multicolumn{6}{c}{global} & \multicolumn{3}{c}{local} \\ \cline{3-8} \cline{9-11}
strain [\%] & stress [GPa] & $K$ [eV~b] & $F_k$ [eV] & $\sigma$ [GPa] & $E^*$ [eV] & $d^*$ [b] & $W_m$ [meV] & $K$ [eV~b]& $F_k$ [eV] & $\sigma$ [GPa]  \\ \hline
0   & 0    & 1.60 & 0.91 & 0    & 1.94 & $\infty$ & 158$\pm$5 & 1.40 & 0.90 & 0\\
0.5 & 0.3  & 1.34 & 0.88 & 0.29 & 1.20 & 4.13     & 148$\pm$5 & 1.30 & 0.86 & 0.27\\
1.0 & 0.61 & 1.12 & 0.83 & 0.57 & 0.89 & 2.66     & 134$\pm$10  & 1.33 & 0.86 & 0.56\\
1.5 & 0.92 & 1.02 & 0.79 & 0.85 & 0.65 & 1.73     & 117$\pm$15  & 1.33 & 0.85 & 0.84\\
\end{tabular}
\end{ruledtabular}
\end{table*}

\begin{figure}
\begin{center}
\includegraphics*[width=8.6cm]{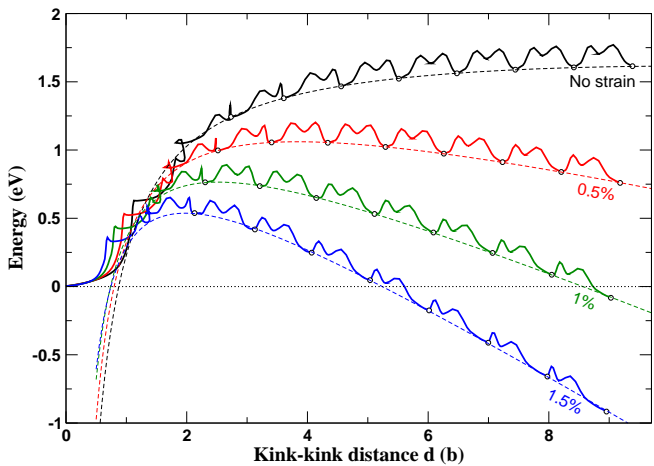}
\caption{(color online) Kink pair energy as a function of kink-kink separation $d$ (in Burgers vector $b$), for several shear strain values. The thick full lines represent the NEB calculations whereas thin dashed lines show the energy variation for stable configurations, as fitted from elasticity theory.} \label{figstress}
\end{center}
\end{figure}

No noticeable changes in the migration and formation mechanisms have been found, for any of the shear values, compared to the no-stress case. However, the energy variation as a function of the kink-kink separation $d$ now shows a different behavior (Figure~\ref{figstress}). After the initial increase due to the formation of the kink pair, the energy now reaches a maximum value, then decreases again when the two kinks get further separated. The maximum is characterized by the energy $E^*$ and the kink-kink distance $d^*$. Both values exhibit a strong dependence on the applied stress: the higher the stress, the lower $E^*$ and $d^*$. 

The elasticity theory predicts an energy variation 

\begin{equation}
\Delta E=-\frac{K}{d}+2\,F_k-\sigma bhd, \label{eqnformigrstress}
\end{equation}

which is similar to the expression~\ref{eqnformigr} but also includes the work done by the applied stress $\sigma$.\cite{Hir82WIL,See06PM} The energy variation as a function of $d$ and stress, using $d$ determined from both the global and the local approachs for stable configurations, has been fitted with this model (Figure~\ref{figstress}). We found an excellent agreement, except for very small $d$, indicating that this simple elastic model correctly describe the kink pair formation in a system under stress. 

The obtained values $K$, $F_k$ and $\sigma$, resulting from the fits, are reported in the table~\ref{stresstable}. $F_k$, which now represents the single kink energy in a stressed system, is decreasing when the applied stress increases. This diminution is modest in the case of the global approach, and even smaller for the local approach. The fitted quantity $\sigma$ should theoretically be equal to the applied stress. Here, although close, the fitted value tends to be systematically lower than the applied stress. A posssible explanation is that a fraction of the applied stress is relaxed through free surfaces having \^X as normal. 

From the NEB calculations, $E^*$ and $d^*$ can be determined as a function of stress (Table~\ref{stresstable}). Those are important quantities since they define the energy barrier to overcome to form a stable kink pair. In a non stressed system, the kink pair is never stable and the asymptotic value $E^*$ is simply $2F_k+W_m=1.94$~eV. Our results indicate that both $E^*$ and $d^*$ are strongly dependent on the stress. For the largest applied stress considered here, about 1~GPa, we found that the energy barrier ($E^*$) is reduced to 0.65~eV and is reached as soon as kinks are separated ($d^*$) by 1.73~b, i.e. 6.64~\AA. 

Finally, the kink migration barriers $W_m$ have been determined for each applied stress, by substracting the energy variations given by eq.~\ref{eqnformigrstress} from the initial energy curves. Computed values, obtained for large kink-kink distances, are reported in the table~\ref{stresstable}, and indicate that $W_m$ is sligthly decreasing when the stress increases. It is also interesting to examine the variation of the energy barriers for migration in the initial energy curves, shown in the figure~\ref{figstress}. The work done by the stress during the kink migration allows to further reduce the energy barrier. For instance, the latter drops to 75~meV for the largest stress considered here.  

\section{DFT results}\label{DFT}

\subsection{Single kink: structure}

\begin{figure}
\begin{center}
\includegraphics*[width=8.6cm]{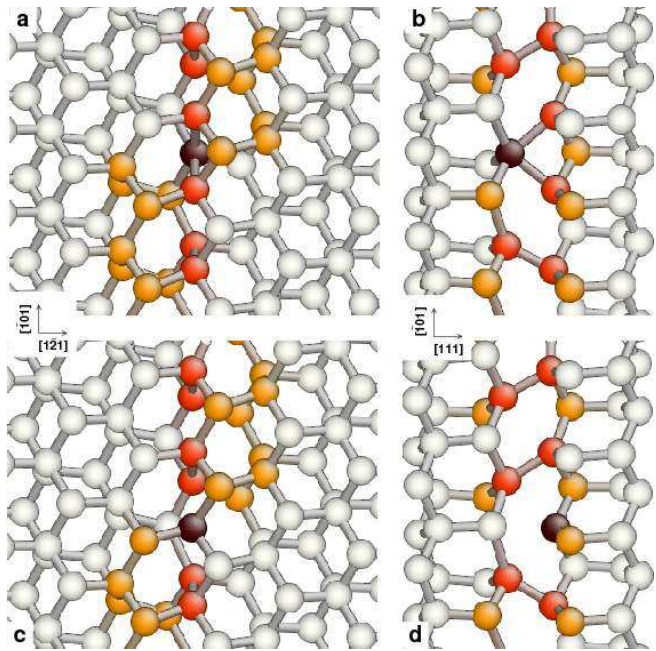}
\caption{(color online) Ball-stick representations of two kink structures stable in DFT calculations (top: narrow kink, bottom: DB kink), along two different orientations, (111) (left) and $(1\bar{2}1)$ (right) projections. Same color convention than in figure~\ref{Kinkgeometry}.} \label{KinkstructDFT}
\end{center}
\end{figure}

We now focus on the results obtained from first-principles calculations, in particular the structure of the kink. The section~\ref{singlekink} described single kink EDIP calculations using a large cell and fixed boundaries. From this structure, a cluster-like system of 529 Si atoms, centered on a relaxed narrow kink, has been cut out. The surface Si atoms were kept at fixed positions and saturated with a total of 271 hydrogen atoms. Forces relaxation yields a configuration close to the initial guess, as shown in the upper part of the figure~\ref{KinkstructDFT}, proving that the narrow kink configuration is stable within DFT. The two bonds in the kink core have a bond length of 2.67~\AA, and are tilted by 39$^\circ$ relatively to $(\bar{1}01)$. For comparison, a straight screw dislocation relaxed within DFT exhibit dimers with a bond length of 2.46~\AA\ and a tilting angle of 21$^\circ$. This suggests that the core of the narrow kink is characterized by largely stretched bonds. Performing EDIP simulations with the same system, we found a bond length of 2.53~\AA, indicating stronger bonds, with a rather similar tilting angle of 39$^\circ$. 

Relaxation from different initial structures revealed one new kink configuration, shown in the lower part of Figure~\ref{KinkstructDFT}. When compared to the narrow kink, a stretched bond in the kink core no longer exists, causing a 3-coordinated atom. We called this configuration the DB (Dangling Bond) kink. The two dimers on both sides on the 3-coordinated atom are characterized by a length of 2.52~\AA, and a tilting angle of 31$^\circ$ relatively to $(\bar{1}01)$ (Fig.~\ref{KinkstructDFT}d). If the center of the kink is defined by its geometry (local determination), it appears that the location of the DB kink is shifted by approximately $b/2$ compared to the narrow kink. 

It would be interesting to know whether the narrow kink or the DB kink is the most stable geometry. Unfortunately, both configurations are very dependent on the initial conditions, i.e. on the stress state imposed by the rigid surface. In fact, it was not possible to obtain the DB kink from the system with surfaces to conform a narrow kink, and vice versa. Also, the calculated total energies cannot be directly compared since they have been obtained with different fixed surface configurations. The kink structure results from a competition between an elastic energy and a chemical energy. For the narrow kink, the gain in chemical energy by forming bonds is balanced by the loss in elastic energy when atoms are moved closer together. For the DB kink, it is the opposite. We will show in the next section that both configurations are quasi degenerate.  

\subsection{Migration}

\begin{figure}
\begin{center}
\includegraphics*[width=8.6cm]{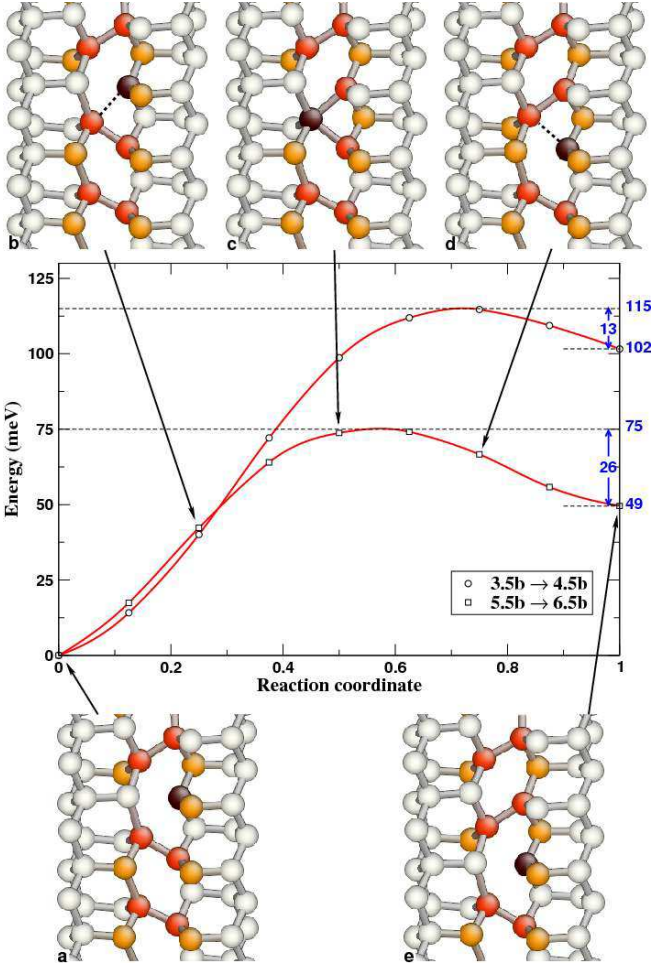}
\caption{(color online)  NEB calculated excess energy versus reaction coordinate (middle graph) corresponding to the migration of one kink, for two different kink-kink separations $d$, and ball-stick representations ($(1\bar{2}1)$ projection) of the successive structures (same color convention than in figure~\ref{Kinkgeometry}) for $d$ ranging from $5.5b$ to $6.5b$. } \label{DFTmigration}
\end{center}
\end{figure}

The kink migration process has been investigated by performing NEB calculations with 9 images. Initial and final states are obtained from large cell configurations relaxed with the EDIP potential and originate from the computations of kink pair energies as a function of the kinks separation. Clusters were cut from these configurations, in such a way that a kink was located in its center. Surface atoms, saturated by hydrogen atoms, are kept fixed at their initial positions to retain the displacement field for two opposite kinks in interaction, as relaxed by the EDIP potential. With this procedure, we are able to mimic the migration of a single kink with an increasing kink-kink separation, using only one single kink in the cluster. The method relies on the two following assumptions: (i) during the kink migration the interaction between the kink and the surface does not change significantly (ii) the EDIP potential correctly describes the deformation field far from the kink. The cluster cells approximate dimensions are  $6\mathbf{i}\times12\mathbf{j}\times5\mathbf{k}$, and it consists of 336 Si and 197 H atoms. 

Three cases have been considered, corresponding to increases in the kink-kink separation $d$ from $1.5b$ to $2.5b$, $3.5b$ to $4.5b$, and $5.5b$ to $6.5b$. In the first case, $d$ is small enough for both kinks to be included in the cell. However, the kinks are too close to be stable and relaxation leads to their annihilation, suggesting that an investigation of the mechanism of the kink pair formation within DFT using the reduced size systems is impossible. The MEPs for the two other cases, as well as intermediate configurations during kink migration, are shown in the figure~\ref{DFTmigration}. 

For both cases, relaxation of the initial and final images leads to a stable DB kink (fig.~\ref{DFTmigration}a,e). Starting from this geometry, the kink migration occurs by formation and breaking of one bond. First, the 3-coordinated atom moves toward the atom defining the next dimer along the migration direction. This dimer is progressively stretched (fig.~\ref{DFTmigration}b), until a bond is formed. At this point, the process is halfway and the structure corresponds to a narrow kink (fig.~\ref{DFTmigration}c). The dimer is further stretched, and finally breaks (fig.~\ref{DFTmigration}d). The DB kink structure is recovered but now shifted by $b$ along $[\bar{1}01]$.  

Figure~\ref{DFTmigration} shows the DFT calculated energy variation associated with the kink migration. For both cases, the curves are simpler than the EDIP results (fig.~\ref{EDIPmigration}), since no intermediate metastable configuration appears. Nevertheless, similar features are observed. In fact, the final configurations have energies larger than initial ones, and migration from $5.5b$ to $6.5b$ is associated with a smaller barrier than migration with $d$ ranging from $3.5b$ to $4.5b$. This being due to the long range elastic interaction between the two kinks. 

For both MEPs, the migration energy is given by the energy maxima, which correspond approximately to a narrow kink structure, with values of 115 and 75~meV. If the migration energy decreases as an inverse power law as a function of $d$, what was shown in the large scale EDIP calculations, it is possible to extrapolate the asymptotic limit. For DFT, we found that this limit is close to zero. As a result, the DB and narrow kink configurations would be quasi degenerate in energy for an isolated kink, explaining why the outcome of single kink relaxations depends on fixed surface conditions. Nevertheless, our kink migration calculations suggest that in presence of another kink, the DB kink configuration is slightly more stable.  

The energy barriers for a reverse kink migration process, i.e. for a decrease of $d$ and a subsequent recombination of the kinks, are indicated in the figure~\ref{DFTmigration}. They are very small, with respective values of 13 and 26~meV, and provides a simple explanation of the spontaneous annihilation of kinks for $d$ smaller than $3b$.  

\subsection{Single kink: energy}

\begin{figure}
\begin{center}
\includegraphics*[width=8.6cm]{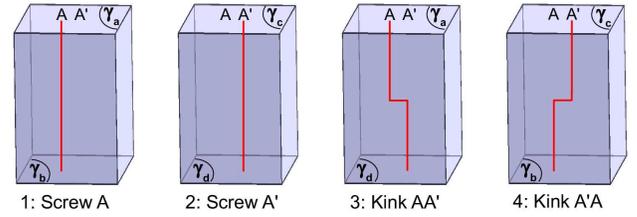}
\caption{(color online) Sketch of the 4 systems used for the determination of the kink formation energy.} \label{DFTenergy}
\end{center}
\end{figure}

Large cell EDIP calculations allowed us to determine the kink formation energy $F_k$ from the energy variation as a function of the distance between kinks. However, since DFT computations implie much smaller cells, $F_k$ has to be computed by other means. The two main issues are the small cell dimensions and the fact that only total energy is accessible from first principles calculations. We used small cluster cells with fixed surfaces, leading to non negligible interactions between screw segments and the surfaces. Also, the energies of the cluster surfaces have to be known. It is likely that both contributions are of the same order or larger than $F_k$. 

Nevertheless, in the following we show that it is possible to determine $F_k$ by combining the energies of four different systems, schematically depicted in the figure~\ref{DFTenergy}. System 1 (2) is a perfect screw dislocation located in A (A') (labelled as in figure~\ref{Kinkgeometry}). System 3 (4) includes a kink changing the screw location from A to A' (from A' to A). All systems are constructed in such a way that the kink is in the center of the cell. The total energies of the 4 systems are 

\begin{subequations}
\begin{eqnarray}
E_1 & = & E_\mathrm{screw}+I_1+\gamma_a+\gamma_b, \\
E_2 & = & E_\mathrm{screw}+I_2+\gamma_c+\gamma_d,\\
E_3 & = & E_\mathrm{screw}+I_3+\gamma_a+\gamma_d+F_k, \\
E_4 & = & E_\mathrm{screw}+I_4+\gamma_c+\gamma_b+F_k.
\end{eqnarray}
\end{subequations}

Here, $E_\mathrm{screw}$ is the screw dislocation energy while $\gamma_{a-d}$ are energies of the \^Z-surfaces (Figure~\ref{DFTenergy}). $I_i$ are interaction terms and include contributions from the interaction between the fixed boundaries, and both the screw segments and the kink. Simple arithmetic leads to the following expression where the surface energies $\gamma_{a-d}$ have cancelled out

\begin{equation}
2F_k=\left[\left(E_3+E_4\right)-\left(E_1+E_2\right)\right]+\Theta \label{eqnDFTenergy}
\end{equation}

with $\Theta=\left[\left(I_3+I_4\right)-\left(I_1+I_2\right)\right]$. Theoretically, $\Theta$ is simply twice the energy contribution coming from the interaction between a single kink and the fixed boundaries, and should be negligible in a large cell. We checked this assumption by computing $\delta E=\left(E_3+E_4\right)-\left(E_1+E_2\right)$ using the EDIP potential and large cells ($21\mathbf{i}\times42\mathbf{j}\times20\mathbf{k}$) with fixed boundaries. We found $\delta E$=1.83~eV, i.e. very close to $2F_k$=1.82~eV as determined in the section~\ref{EDIP}. 

Initial structures for first-principles calculations have been prepared by cutting out clusters, centered on kinks, from large systems relaxed with the EDIP potential.  This procedure was used to reduce the unwanted interactions between the kink and the surfaces. Furthermore surface atoms with three dangling bonds were removed. All remaining surface atoms have been locked to their original positions, and saturated by hydrogen atoms. The clusters include 529~Si+270~H (system (1)), 528~Si+270~H (system (2)), 529~Si+271~H (system (3)) 528~Si+269~H (system (4)), resulting in dimensions close to $6\mathbf{i}\times11\mathbf{j}\times8\mathbf{k}$. One can see that the total numbers of Si and H atoms in (3)+(4) and (1)+(2) are equal, validating equation~\ref{eqnDFTenergy}. 

Forces relaxation for systems (3) and (4) leads to different kink configurations, a narrow kink structure (3) and a DB kink (4). However, we have already shown in the previous section that these two kink structures are almost degenerate in energy. Using equation~\ref{eqnDFTenergy}, our calculated $\delta E$ is 2.78~eV. However, it is likely that $\Theta$ is not negligible for the small systems employed in DFT calculations. To estimate this term, EDIP calculations with equivalent systems were performed. We found $\delta E$=1.89~eV, giving $\Theta$=0.07~eV, provided that the EDIP kink formation energy is $F_k$=0.91~eV. Using this value in Eq.\ref{eqnDFTenergy}, we found the DFT kink formation energy to be $F_k$=1.36~eV. 

\section{Discussion}

\subsection{Kink structure and stability}

Simulations have revealed that three kink geometries are possible on an undissociated screw dislocation. Two kink configurations, labelled narrow and wide, are stable when using the EDIP potential (fig.~\ref{EDIPkinkstruct}). For DFT calculations, we found two stable configurations, the same narrow structure and a new one, labelled DB (fig.~\ref{KinkstructDFT}). Actually, the three possible configurations are closely related. A kink on a screw dislocation line is characterized by the necessary reversing of the tilting angle of dimers stacked along the axis \^Z at the B position (see Fig.~\ref{Kinkgeometry}). The three kink configurations possess this feature, at the expense of one (narrow kink) or three (wide kink) 5-coordinated atoms, or one 3-coordinated atom (DB kink). All attempts to obtain other stable geometries, such as with no over- or sub-coordinated atoms, were unsuccessful. This is an entirely different situation compared to partial dislocations in silicon,\cite{Nun00JPCM,Bul95PMA,Bul01PMA} or to screw dislocations in bcc metals,\cite{Due83AME,Due83AME2,Yan01MSE,Wan03PRB} for which a large number of different kink geometries are possible. 

The investigation of the kinks stability indicates that for EDIP the narrow kink is favored, while for DFT the narrow and DB kink are degenerate in energy. Additional EDIP calculations showed that for EDIP an initial DB kink structure relaxes to a narrow kink. We did not try to calculate a wide kink with DFT, since it is very unlikely that it would be stable. Hence, two stable configurations are possible for a kink on a perfect screw dislocation. In view of the reduced sizes of our systems, limiting the accuracy of the calculations, it would not be credible to assert which of the configuration that is the more stable one. On the basis of DFT calculations, it is reasonable to conclude that such an information is not relevant, as the energy difference is so small that a kink would not stay in one single defined geometry for a significant amount of time. 

\subsection{Kink migration}

An energy barrier of 158~meV for the kink migration process is calculated with EDIP, whereas it is found to be a negligible barrier within DFT. Even if the latter is likely associated with uncertainties due to the reduced sizes of the computational systems, it appears that the migration barrier is nevertheless very small. Such a small energy is common for metals, but not for covalent materials like silicon. Then, at room temperature, kinks along the screw dislocation in silicon should be highly mobile. 

EDIP and DFT calculations lead to different structural changes during migration. However, the mechanisms allowing the kink movement are similar in both cases, since it requires the formation and the breaking of one single bond. These bonds are highly stretched, in particular in DFT, suggesting that the energy cost for breaking or forming a bond is low, what explains the low kink migration energy determined with both methods. 

\subsection{Kink formation}

\begin{figure}
\begin{center}
\includegraphics*[width=8.6cm]{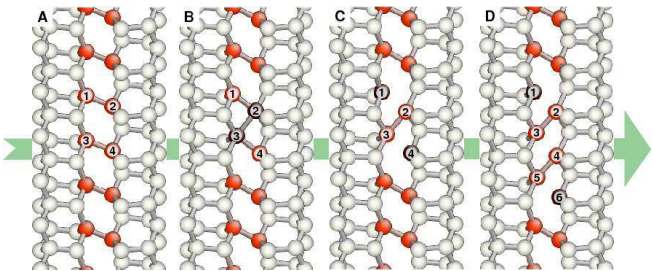}
\caption{(color online)  Ball-stick representations ($(1\bar{2}1)$ projections) of possible successive structures leading to a kink pair formation from a straight screw dislocation. Same color convention than in figure~\ref{Kinkgeometry}.} \label{DFTformation}
\end{center}
\end{figure}

We have proposed a possible mechanism for the formation of a kink pair along a perfect screw dislocation, on the basis of EDIP calculations (Fig.~\ref{EDIPformation}). Like all mechanisms described in this work, it requires the formation and breaking of bonds between atoms belonging to dimers stacked along \^Z. A single kink is characterized by a reversing of the dimers tilt. The formation of a kink pair then requires two changes in the tilt of the dimers, what also can be observed in figure~\ref{EDIPformation}c. The creation of the kink pair is associated with a significant energy increase. After formation, the kinks still have to overcome a large elastic interaction in order to separate from each other. Reduced system sizes and the very small recombination barrier did not allow us to investigate kink pair formation within DFT. Nevertheless, we propose a possible process, depicted in the figure~\ref{DFTformation}. First steps are similar to the EDIP result, with the formation of one bond between atoms 2 and 3 (Fig.~\ref{DFTformation}b). Then, unlike the case for EDIP, bonds 1-2 and 3-4 break, leaving only one single dimer with a reverse tilt and two 3-coordinated atoms (Fig.~\ref{DFTformation}c). This geometry, simple and very narrow, includes a pair of opposite kinks, although it is almost certain that it is not stable. Subsequent kink migration leading to an increasing separation would, however, stabilize the kink pair (figure~\ref{DFTformation}d).

A single kink formation energy $F_k=0.90$~eV has been determined for the EDIP potential from the creation and separation of a kink pair. Using DFT, we have computed  that $F_k=1.36$~eV by combining the results of four total energy calculations. It is rather difficult to propose one unique final value from the two models. While EDIP computations have been performed in large systems, with a fine control of the effect of boundary conditions, semi-empirical potentials for silicon often suffer from an intrinsic lack of quantitative accuracy. Conversely, a better determination is expected with DFT, but in that case, boundary effects could be much larger and difficult to estimate, because of the reduced sizes of computational systems. Due to surface constraints, it is likely that the DFT result is an upper limit of the true value, that is why the kink formation energy should be within the range from 0.9 to 1.36~eV. 

\subsection{Effect of stress}\label{stressdiscuss}

\begin{figure}
\begin{center}
\includegraphics*[width=8.6cm]{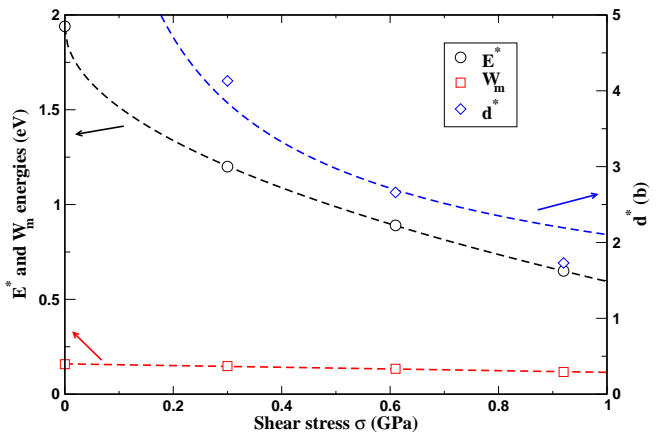}
\caption{(color online) Variation of $E^*$ (circles), $d^*$ (diamonds), and $W_m$ (squares) as a function of the applied stress. Dashed lines are fit from the formulas described in the section~\ref{stressdiscuss}.} \label{varstress}
\end{center}
\end{figure}

Calculations with an applied shear stress have been only performed with the EDIP potential. The results, reported in the figure~\ref{figstress}, can be easily fitted using equation~\ref{eqnformigrstress}. This suggests that despite the short separation, the kink energy is well described by the elastic interaction model.\cite{Hir82WIL,See06PM}  According to this model, the saddle point for the kink pair stability is defined by 

\begin{equation}
E^* =  2\left[F_k-(K\sigma bh)^\frac{1}{2}\right],
\end{equation}
and
\begin{equation}
d^* = \left(\frac{K}{\sigma bh}\right)^\frac{1}{2}.
\end{equation}

These expressions have been used for to fit the data reported in the table~\ref{stresstable} and shown in Figure~\ref{varstress}. It appears that the energy barrier $E^*$ variation as a function of shear stress is accurately described by the elastic interaction model. The best agreement is obtained for $F_k=0.97$~eV and $K=1.47$~eV~b. Using these parameters, it can be estimated that the barrier for the formation of a stable kink pair would vanish for an applied stress of $\sigma\simeq2.1$~GPa. It is more difficult to fit the variation of $d^*$. A possible explanation is that it is difficult to determine precisely the distance beetween the two kinks when they are close. 

Finally, the figure~\ref{varstress} also shows the variation of $W_m$ as a function of the applied stress. The best fit is obtained with a linear regression and a coefficient of 45~meV~GPa$^{-1}$. Such a variation indicates that a kink on the screw dislocation cannot be strictly compared to a small edge segment. In fact, otherwise, the force due to the applied shear stress would be zero, and $W_m$ would remain constant. However, it is difficult to make assertions here because of the limited amount of data and the very low migration energies.  

\subsection{Other zinc blende materials}

All our investigations concerned silicon, because of the large body of experimental and theoretical data. However, it is interesting to discuss how our results could be appropriate for other materials with zinc-blende structure. For example, germanium has been shown to have similarities with silicon regarding dislocation properties. In the low temperature/large stress regime, screw dislocations are also expected to be non-dissociated and located in the shuffle set.\cite{Piz05EPL} It is then likely that the kinks creation and migration mechanisms proposed in this paper could also be valid for Ge. If so the kink migration barrier could be approximated as the energy difference between the DB and the narrow kinks, or between the narrow and the wide kinks. It is however difficult to determine which of the three possible kink structures will be the more stable one. While it is less energetically expensive to have overcoordinated atoms in Ge (compared to Si) thus favoring the narrow and the wide kinks, the energy penalty of a dangling bonds is also smaller in Ge, thus favoring the DB kink. It is likely that both effects are of the same order, leading to a very small kink migration energy, as in silicon. Of more interest is the kink formation energy. First principles calculations of the Peierls stress of a non-dissociated screw dislocation reveal a value approximately four times smaller for Ge than for Si.\cite{PizUNP} Assuming that the kink formation process is directly related to the Peierls stress, a single kink formation energy should be as low as 0.34~eV in Ge. 

Cubic silicon carbide is another zinc-blende semiconductor for which we can extrapolate our findings. Previous calculations have shown that glide and shuffle screw dislocations were very close in energy,\cite{Piz05EPL} in agreement with experiments indicating the coexistence of shuffle perfects and glide partials.\cite{Dem05PSS} We focus on the kink pairs formation and migration for perfect screw dislocation in the shuffle set. Kink structures in silicon carbide could be more complex than in silicon due to the fact that SiC is a binary compound. Indeed, on a non dissociated screw dislocation, a formed kink pair includes now two non-symmetric kinks, with two different migration energy barriers. It is also likely that the kink pairs formation is more complex that what we found for silicon. Nevertheless, if we follow arguments developed above for Ge, we can predict that the kink migration energy is also negligible in silicon carbide. Also, other calculations revealed that the SiC Peierls stress is nearly twice that in silicon,\cite{PizUNP} why it is likely that the kink formation energy is also about twice as the one in silicon. 

\subsection{Comparison with partials}

Previous studies have determined kink pairs formation and kink migration energies for silicon partial dislocations. The situation is more complex than for perfect dislocations, since there are many possible kink configurations, with close energies. Experimentally, Kolar et al have measured formation energies $F_k(90^\circ)=0.73$~eV and $F_k(30^\circ)=0.80$~eV, and a kink migration energy of $W_m=1.24$~eV.\cite{Kol96PRL} Theoretically, consensus has not been obtained: formation energies range from 0.04~eV to 1.2~eV for 90$^\circ$, and from 0.25~eV to 2.15~eV  for 30$^\circ$, migration energies from 0.6~eV to 1.8~eV  for 90$^\circ$, and from 0.7~eV to 2.1~eV for  30$^\circ$.\cite{Leh99EMIS,Bul95PMA,Bul01PMA,Hua95PRL,Obe95PRB,Val98PRL} Here, we consider the experimental data for partials, that we compare with our first principles computed values. 

Our calculated $F_k$ for perfect dislocations is larger than $F_k$ for partial dislocations. Higher values were indeed expected for perfect dislocations, since isotropic elasticity theory\cite{Hir82WIL} indicates that a single kink self-energy is proportional to $b^2$. Since $b(\textrm{screw})\,/\,b(\textrm{partial})=\sqrt{3}$, it is seen that the kink formation energy is three times larger for perfect than for partial dislocations. The here found ratio is smaller, but of the same order. Obviously, elasticity theory only cannot be used for a quantitative determination. 

We found a striking difference between partial and perfect dislocations regarding kink migration. While a value of 1.24~eV has been measured for partial dislocations,\cite{Kol96PRL} we found that in the case of perfect dislocation, the migration energy is extremely low, almost zero. It is difficult to find a simple explanation for such an impressive difference, as the atomistic mechanisms for kink migration are different for each partial dislocations, and also different for the screw dislocation. Nevertheless, it is noteworthy that both the core structures of the screw dislocation and its kinks are free of complex reconstructions. A kink then migrates easily by breaking and forming highly stretched bonds. Conversely, partial dislocations and its numerous possible kinks structures have reconstructed cores, and mechanisms such as stretching, rotation and even breaking of strong bonds are involved in the migration process,\cite{Bul01PMA} thus requiring much larger energies. 

\subsection{Comparison with experiments}

Scratch tests and deformation experiments performed under high confinement pressure revealed that the plasticity of silicon is governed by perfect undissociated dislocations, supposed to be located in the shuffle set. These experiments have been done for temperatures ranging between 150K and 700K.\cite{Rab00JPCM,Rab00PSS,Rab01SM,Asa05MSE} Recently, using two different methods, Rabier et al. have determined that local stresses required for displacing these perfect dislocations are larger than 1~GPa at  573K.\cite{Rab05PSSb,Rab07PSSb} Then, at lower temperatures (i.e. well below the brittle-ductile transition temperature), very large stresses are required in order to move the  dislocations.  

These data are in very good agreement with our model for non-dissociated dislocations, located in the shuffle set, and moving via the thermally activated formation and migration of kinks under a large stress. In fact, for an applied stress of about 1~GPa, we found that the energy barrier for the formation of a stable kink pair is as low as 0.65~eV (Table~\ref{stresstable}). Considering that the dislocation mobility could be described using transition state theory, it is possible to determine the average time required for the formation of one stable kink pair as a function of temperature from the expression

\begin{equation}
t = \frac{1}{\nu_0}\exp\left(\frac{E^*}{kT}\right), 
\end{equation}

neglecting entropic contributions. Using $\nu_0\simeq10^{13}$~Hz, the silicon Debye frequency,  and $E^*=0.65$~eV, we found $t=8$~ms at 300K. Consequently, perfect dislocations could move easily via a kink mechanism at room temperature for applied stresses of the order of a 1~GPa. 

\section{Conclusion}

First principles and interatomic potential calculations have been performed to investigate the formation and migration of kinks on a non-dissociated screw dislocation in silicon. We found that the structure of a single kink can be characterized by an angular switch of the tilt for dimers stacked along $[\bar{1}01]$. Two or three configurations, being very similar in geometry, are nearly degenerate in energy. The single kink energy has been computed to range between 0.9~eV and 1.36~eV. The kink migrates by successively alternating between two configurations, resulting in a very low migration energy, below the uncertainly limit. As a consequence, when formed, kinks on a non-dissociated screw dislocation in silicon will migrate freely at room temperature. We have also investigated the kink pair formation mechanism and its energetics using an interatomic potential, and how an applied stress modifies the formation and migration process. We found a strong stress effect, an applied stress of about 1~GPa leading to a 66\% reduction of the initial energy barrier for formation. These results are in very good agreement with experiments on silicon performed in the low temperature/high stress regime. 

There have been several investigations of the properties of kinks on partial dislocations in silicon. Gathering the large compilation of data with our results, it would be interesting to redo the analysis of Duesbery and Jo{\'o}s about the glide/shuffle controversy.\cite{Due96PML} Such a study would further require a determination of the energy barriers associated with the cross-slip from the shuffle to the glide plane, and the dissociation into partial dislocations, what has not been done to our knowledge. Nevertheless, such an investigation would make it possible to understand and characterize the transition between the high temperature/low stress and temperature/high stress regimes. Future researches should be oriented in this direction. 

\section*{Acknowledgements}

One of us (L.P.) thanks the Poitou-Charentes region and the University of Iceland for financial support during his stay in Reykjavik. This work was supported by the SIMDIM project under contract N$^\circ$~ANR-06-BLAN-250.

\end{document}